\documentclass[aps,prb,twocolumn,floats,showpacs]{revtex4}
\usepackage{graphicx}
\usepackage{color}
\usepackage{bm}
\newcommand{\im}{\mbox{Im}}

\begin{document}
\title{Shot noise in Graphene with long range Coulomb interaction and the local Fermi distribution}
\author{  Anatoly Golub and  Baruch Horovitz}
\affiliation{ Department of Physics, Ben-Gurion University, Beer Sheva 84105 Israel\\   }
\begin{abstract}
We calculate the shot noise power in ballistic graphene using the kinetic equation  approach based on the Keldysh technique. We find that the local energy distribution function obeys Poisson's equation, indicating a mapping into a diffusive metal system. We derive the conductance and noise including the long range Coulomb interaction to first order. We find that the shot noise increases due to interaction, leading to a frequency dependence. Furthermore, we find that the Fano factor at degeneracy is 1/3, the same as without the Coulomb interaction.
\end{abstract}
\maketitle
\section
{ Introduction}
Graphene, a two-dimensional honeycomb lattice of carbon, is of considerable recent interest. It is now experimentally established \cite{geim,guinea} that a great deal of properties of graphene can be understood in terms of noninteracting linear dispersing Dirac quasiparticles. A particular attention focuses on ballistic transport. To identify the elusive long range Coulomb interaction we study here shot noise, which is a fundamental nonequilibrium quantity.

Conductivity and shot noise $S_n$ have been experimentally measured in graphene \cite{miao,markus,hakonen}. The conductivity for a wide graphene ribbon (the  width of the ribbon exceeds the its length: $W>>L$) was predicted \cite{kat,been,ludwig1} to exhibit a minimum $\sigma_{min}$ at the Dirac degeneracy point, while the Fano factor $F=S_n/(2e\sigma_{min}V)$ had a maximum close to 1/3 ($V$ is the voltage).  The conductivity near the neutrality point is attributed to the fact that the current
is mediated by evanescent rather  than propagating
modes. The transport by evanescent  channel is suppressed by a factor $\exp(-k_y L)$ (here $k_y$ is the momentum in the width $W$ direction). However, in the case $W>>L$
  the exponent is actually small ($k_y L<<1$) for many channels, resulting in an universal minimal value for conductivity at the Dirac point.

   The minimum conductivity
at the graphene Dirac point can be reached in clean disorder-free graphene. To achieve a such regime may be a problem even for ultraclean high mobility suspended graphene \cite{bolotin} with a possibility of inhomogenous  charge landscape \cite{sarma}due to charged impurities. The measurements \cite{geim,bolotin} result
in a conductance near neutrality point larger than theoretical values, while other data \cite{hakonen} do obtain $\sigma_{min}$. The origin of this discrepancy is unclear at present.

The Fano factor has  been analyzed theoretically in a number of works \cite{been,mirlin,ludwig} relying on  the Landauer approach using a distribution of transmission eigenvalues; this approach is limited to non-interacting systems. At low temperatures a universal minimum conductivity is found
$\sigma_{min} = (4/\pi)e^2/h$ as well as a universal Fano factor
 $F= S/2eI = 1/3$.
Interestingly, the $F=1/3$ result is also known in diffusive metals.

The Landauer approach is difficult to generalize when electron-electron interactions are present. Here we apply the kinetic equation method, which is based on the Keldysh technique. This method is an alternative that permits to handle the Coulomb interaction near the Dirac point. A second advantage of our method is that the energy distribution function can be readily identified.
We address zero-frequency shot noise; however, at low $\omega$ a frequency dependence of the shot noise will be included through the Coulomb coupling renormalization.

Our results can be tested by two specific experiments: first is the probe of local energy distribution as function of the distance $R$ from one of the electrodes, as done e.g. in disordered wires \cite{pothier}. We predict that the distribution has two steps and it interpolates linearly with $R$ between the Fermi distributions with voltages $0,V$ of the two electrodes, respectively. A second experiment tests the presence of the Coulomb coupling $g$ by looking at the conductivity or at the shot noise at finite frequency $\omega$. The correction to the noise varies [Eq. (\ref{ST}) below] as $g_r\sim 1/\ln(\omega)$ where $g_r$ is the renormalized \cite{guinea3,herb} interaction at finite $\omega$.

The paper is organized as follow. In Sec. II we introduce the Keldysh action of the system and present the formula of noise as a variation in the partition function with respect to the quantum components of the vector potential. In Sec. III we consider the noninteracting ballistic graphene near the Dirac neutrality point. For this purpose we develop a new approach based  on a quasiclassical approximation: by gradient expansion we obtain the kinetic equation for the non-equilibrium distribution function. This kinetic equation
 is a diffusion type similar to the one which describes one dimensional dirty wires. In Sec. IV we consider the impact of long range Coulomb interaction and present the results for conductance and shot noise power. The main calculations which are related to this section are given by  Appendices (A and B).

\section { Keldysh action} We use the standard form for the Hamiltonian that describes graphene, being equivalent to QED in 2+1 dimension \cite{guinea,gus}
 \begin{equation}
 {\cal H}=\sum_{\sigma}\bar{\psi}_{\sigma}[i\gamma^x\hbar v_F(\partial_x+\frac{ie}{\hbar c}\bar{A}_x)+i\gamma^y\hbar v_F\partial_y]{\psi}_{\sigma}
\end{equation}
 Here $\psi_{\sigma}$ are 4 component spinors corresponding to two inequivalent Dirac points and to two atoms in the unit cell, $\sigma$ is a spin index and $\bar{A}_x$ is the $x$ component of the vector potential. The Dirac matrices satisfy the standard algebra $\{\gamma_{\mu},\gamma_{\nu}\}=2g_{\mu\nu}$
with the explicit representation  \cite{gus}
 \[ \gamma_0=\left(
                                        \begin{array}{cc}
                                          0 & I \\
                                          I & 0 \\
                                        \end{array}
                                      \right),\,\,\,\ \vec{\gamma}=\left(
                                                               \begin{array}{cc}
                                                                 0 & -{\bm \sigma} \\
                                                                 {\bm \sigma} & 0 \\
                                                               \end{array}
                                                             \right)\]
where $\vec{\sigma}$ are Pauli matrices.

 We write the corresponding action in the Keldysh rotated form. We consider the multichannel limit when the width $W$ of the graphene layer is bigger than its length $L$, i.e. the aspect ratio is  $W/L\gg 1$. This allows us to replace the summation over channels by integration over momentum. The graphene layer occupies the space $0<x<L$, and we consider the fluctuations of the total current (integrated over $W$), therefore only the $x$-component of the electromagnetic field is needed. Thus, the total action as a function of the vector potential (its classic and quantum components) acquires a form $S(A)=S_0 (A)+S_{int}$
\begin{eqnarray}
S_0(A)&=&\sum_{\sigma }\int
d^2 x dt\bar{\psi}_{\sigma}(\hat{G}_{\sigma}^{-1}-\frac{ev_F}{c}\gamma_x \hat{A}_x){\psi}_{\sigma}
 \label{S0}
\end{eqnarray}
where $\hat{A}_x=A_x\tau_0 +\tau_x A^q _x$;  $A_x$ and $A^q _x$ are the classical
and quantum components of the vector potential, respectively \cite{kamenev}.
Here $\tau_x $ is a Pauli matrix and $\tau_0$ is the unit matrix, both act in the Keldysh space, and ${\psi}_{\sigma}$ becomes an 8-component spinor including the Keldysh indices. The Green's function has a form
\begin{equation}
\label{prime}
    G_\sigma = \left(\begin{array}{cc}
 G_\sigma ^R & G_\sigma ^K \\
         0 & G_\sigma ^A
\end{array}\right )
\end{equation}
with each entry as a matrix in Dirac space. The classical field $A_x$ is included in $G^{-1}$, hence
\begin{equation}\label{inv}
    (G_\sigma ^R)^{-1}(A) = i\gamma^0\hbar\partial_t +
    i\gamma^x\hbar v_F(\partial_x+\frac{ie}{\hbar c}A_x)+i\gamma^y\hbar v_F\partial_y
\end{equation}

The unscreened long range Coulomb interacting is given by the part of the action which in the rotated Keldysh basic acquires a form(we set $v_F=\hbar=1$)
\begin{eqnarray}
S_{int}&=&\frac{g}{2}\sum_{\sigma, \sigma' }\int d^2 x d^2 x' dt
(\bar{\psi}_{\sigma}({\bf x}t)\gamma^0 \tau_0 {\psi}_{\sigma}({\bf x}t))\frac{1}{|{\bf x}-{\bf x}'|} \nonumber\\
&&(\bar{\psi}_{\sigma'}({\bf x}'t)\gamma^0 \tau_x {\psi}_{\sigma'}({\bf x}'t))
 \label{Sint}
\end{eqnarray}

 To first order in the interaction the partition function
 $Z(A)=\int D(\bar{\psi}\psi)\exp(i S(A))$ becomes
$
    Z(A)= Z_0 (A)[1+i<S_{int}>(A)]$, where $Z_0 (A)$ corresponds to the action $S_0 (A)$.

The current-current correlation function can be obtained by taking the second derivative of the
source-dependent partition function with respect to the quantum component of the vector potential:
\begin{eqnarray}
S_{n}(t,t')&=& \int\frac{dR}{2L}\int d^2r\frac{\delta^2 \ln{Z(A)} }{\delta A^q_x (\textbf{x}t)
\delta A^q_x (\textbf{x}'t')}|_{A^q \rightarrow0}
 \label{sn}
\end{eqnarray}
here $R=(x+x')/2$ and ${\bf r}={\bf x}-{\bf x}'$.
The leads serve as reservoirs of equilibrium electrons, and we take
 two arbitrary sections $x, x'$ of the graphene area. A similar approach was
 undertaken for calculation of the shot noise
 in dirty wires \cite{levitov}.

\section { Noninteracting ballistic graphene}  In this case
 the noise power acquires the form
\begin{eqnarray}
S_{n0}(t,t')&=& \alpha \frac{e^2 }{2}\sum_{\sigma }\int\frac{dR}{L}\int dr
Tr[G_{\sigma}({\bf x}t{\bf x}'t')\nonumber\\
&&\gamma_x \tau_x G_{\sigma}({\bf x}'t'{\bf x}t)\gamma_x\tau_x]
 \label{sn0}
\end{eqnarray}
where a  $\alpha=W/L$ is the aspect ratio.
In Eq.(\ref{sn0}) we perform the trace in  the Keldysh space
\begin{eqnarray}
S_{n0}(t,t')&=& \alpha \frac{e^2 }{2}\sum_{\sigma }\int\frac{dR}{L}\int dr
tr[G^R_{\sigma}({\bf x}t{\bf x}'t')\gamma_x G^A_{\sigma}({\bf x}'t'{\bf x}t)\nonumber\\
&&+G^A_{\sigma}({\bf x}t{\bf x}'t')\gamma_x G^R_{\sigma}({\bf x}'t'{\bf x}t)\nonumber\\
&&+G^K_{\sigma}({\bf x}t{\bf x}'t')\gamma_x G^K_{\sigma}({\bf x}'t'{\bf x}t)]\gamma_x
 \label{sn00}
\end{eqnarray}
This involves  Keldysh Green's function  which has the standard parametrization
\begin{equation}
G^K=G^R \bar{F}-\bar{F}G^A \label{GK}
\end{equation}
and satisfies Dyson's
equation. The matrix function $\bar{F}$ is the nonequilibrium distribution function \cite{kamenev}. In equilibrium
($V=0$) the  Fourier-transform of this  function is $f_0(\epsilon)=\tanh(\epsilon/2T)$ (for fermions). For graphene with the Dirac hamiltonian, $\bar{F}$ is block diagonal matrix with the two blocks $\bar{F}_- $ and $\bar{F}_+$, each a 2x2 matrix. However, the diagonal elements are dominant at the Dirac point since the off diagonal elements are proportional to the energy deviation from the degeneracy point, i.e. vanish at the Dirac point.
For the ballistic transport regime Dyson's equation is reduced to an equation for the matrix function
$\bar{F}$ which further can be transformed by performing gradient expansion into a kinetic equation. In the clean limit
we have collision-ness limit for this kinetic equation
\begin{equation}
\bar{F}(G^A)^{-1}(A)-(G^R)^{-1}(A)\bar{F}=0
\label{kin}
\end{equation}

The vector potential $A_x$ can be eliminated from this equation by a gauge transformation
\begin{equation}
\bar{F}(\textbf{x}t,\textbf{x}'t') =U(\textbf{x}t)F(\textbf{x}t,\textbf{x}'t')U^{\dagger}(\textbf{x}'t')
\label{U}
\end{equation}
where $U(\textbf{x}t)=\exp[ie\int_0^{{\bf x}} dx'A_x(\textbf{x}'t)]$
and the block entries for $F$ are then $F_- $ and $F_+$.
However, the boundary conditions for $F(xt,x't')$ will be modified to include the
phase factor \cite{levitov}.
By a standard procedure \cite{lifshitz} we obtain the first kinetic matrix equation
\begin{eqnarray}
&&[(-\frac{i}{2}\frac{\partial }{\partial \tau}+\epsilon)F_-
+(\frac{i}{2}\frac{\partial}{\partial R} +p_x)F_-\sigma_x +p_yF_-\sigma_y]-\nonumber\\
&& [(\frac{i}{2}\frac{\partial }{\partial \tau}+\epsilon)F_+
+\sigma_x(-\frac{i}{2}\frac{\partial}{\partial R} +\nonumber \\
 &&+p_x)F_+ +p_y\sigma_yF_+]=0 \label{F1}
 \end{eqnarray}
 Here $\tau=(t+t')/2$ and Fourier transform on the differences
 $ {\bf x}-{\bf x}'$, $ t-t'$  has introduced energy and momentum variables $\epsilon, p_x, p_y$.
The second equation follows from Eq.(\ref{F1}) by replacing $F_{-,+} \rightarrow F_{+,-}$ and $\sigma_x ,\sigma_y \rightarrow -\sigma_x , -\sigma_y$.

We can solve the kinetic equations independently for the $F_{\pm}$ block functions. It is easy to
recover the relation $F_{-}(\epsilon)=F_{+}(-\epsilon)$. Remarkably, we obtain a simple equation
for diagonal function $F_d=(F_-^{11}+F_-^{22})/2$ which is, as we mentioned, of principal importance for noise calculation near degenerate Dirac point
$(\epsilon\rightarrow0)$
 \begin{eqnarray}
\frac{d^2F_d}{d R^2} +k^2 F_d=0
\label{y}
\end{eqnarray}
 where $k=2\epsilon\sqrt{1+p_y^2/(p_x^2-\epsilon^2)}$. Recalling that the boundary conditions for $F$
 include the phase factor we have at the boundaries
 \begin{eqnarray}
F_d(R=0,tt')&=&f_0(t-t')\nonumber\\
F_d(R=L,t,t')&=&f_0(t-t')\exp(-i\phi(t)+i\phi(t'))\nonumber\\
\end{eqnarray}
where the Fourier-transform of $f_0(t-t')$ is $f_0(\epsilon)$ and the phase
$\phi(t)=\int_0^L dx'A_x(x't)$.
For the constant applied bias V we can write $F_d(R,t,t')=F_d(R,t-t')$.  Focusing on the Dirac point,
Eq.(\ref{y}) is reduced to $\frac{d^2F_d}{d R^2}=0$. The solution of this equation is
 \begin{eqnarray}
\frac{F_d(R,t,t')}{f_0(t-t')}&=&1-\frac{R}{L}+\frac{R}{L}\exp(-i\phi(t)+i\phi(t'))\nonumber\\
\label{so2}
\end{eqnarray}
or in Fourier transform
\begin{equation}\label{fermi}
F_d(R,\epsilon)=(1-\frac{R}{L})f_0(\epsilon)+\frac{R}{L}f_0(\epsilon-eV)
\end{equation}
This two step distribution can be tested experimentally at intermediate positions $0<R<L$

In the next step we apply the transformation {Eq.(\ref{U})] to the each Green's function (GF)
in the formula for the noise power (\ref{sn0}), use representation (\ref{GK}) and notice
that the energy integration over the product of GF  of types $G^AG^A$ or $G^RG^R$ gives a zero result.
 The traces  of retarded and advanced GF in the formula for noise are independent of $R$.
 Therefore, we can perform a direct integration in Eq.(\ref{sn0}). The zero frequency noise becomes
\begin{eqnarray}
S_{n0}&=& \alpha\int d\epsilon\kappa(\epsilon)
[1-f_0^2(\epsilon)+\frac{1}{3}(2f_0^2(\epsilon)-\nonumber\\
&&-f_0^2(\epsilon-eV)-f_0(\epsilon)f_0(\epsilon-eV))]
 \label{sn0r}
\end{eqnarray}
where we define
\begin{equation}
\kappa(\epsilon)=\frac{4e^2 }{2}\sum_{\sigma }\int \frac{d^2 p}{(2\pi)^3}
tr(\im G^R_{\sigma}({\bf p}\epsilon)\gamma_x \im G^R_{\sigma}({\bf p}\epsilon)\gamma_x)\,.
\label{con}
\end{equation}
The retarded GF has a form \[G^R_{\sigma}({\bf p}\epsilon)=\frac{(\epsilon+i\delta)
\gamma_0-{\bm \gamma}\cdot {\bf p}}{(\epsilon+i\delta)^2-p^2}\] hence near the Dirac point $\im G^R_{\sigma}(\vec{p}\epsilon)\approx-\gamma_0\delta(p^2+\delta^2)^{-1}$ and
we get (in standard units) $\kappa(\epsilon)=\sigma_{min}$.

Let us consider two limits of Eq. (\ref{sn0r}), the first is the equilibrium noise $T>eV\rightarrow0$,
\begin{eqnarray}
S_{n0}&=& \alpha\int d\epsilon\kappa(\epsilon)
[1-f_0^2(\epsilon)]=4T\alpha\sigma_{min}\,.
\end{eqnarray}
The shot noise corresponds to the limit $V>T\rightarrow0$:
\begin{eqnarray}
S_{n0}&=& \frac{\alpha}{3}\int d\epsilon\kappa(\epsilon)
f_0(\epsilon)(f_0(\epsilon)-f_0(\epsilon-eV))
\end{eqnarray}
Completing the integration we get $S_{n0}=2e|V|\alpha\sigma_{min}/3$, hence the Fano factor is $F=1/3$.

\section { Coulomb interaction} The first order contribution to the noise due to Coulomb interaction (\ref{Sint}) is presented by two topologically different sets of diagrams: one (Fig.1a) is the self-energy contribution, the other (Fig.1b)  is the vertex type diagram. We write them in terms of matrix GF's (in Dirac and Keldysh space).
\begin{figure}
\begin{center}
\includegraphics [width=0.4 \textwidth ]{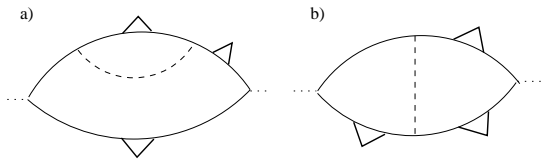}
\caption {A set of diagrams describing the first order corrections to the noise. The dotted lines represents the current vertex, the solid lines stand for retarded or advanced Green's functions, the solid lines with a triangle represent Keldysh functions, and the dashed line refers to the Coulomb interaction. An additional set of diagrams  involves only one Keldysh and three retarded or advanced Green's functions.}
\end{center}
\end{figure}
The detail calculations of these diagram are given in Appendix A. Here we presents the result.
In the case of equilibrium noise Coulomb contribution is
\begin{eqnarray}
S_g&=&0.76\alpha T \sigma_{min}g_r\nonumber
\end{eqnarray}
We note that this value of noise comes out of the nondivergent part of vertex diagram (Fig.1b) while logarithmic divergences in both self-energy and vertex diagrams cancel.

Here we introduce the Coulomb renormalized coupling $g_r$ that
 defines the interaction contribution to the noise. It is known to decrease as a function of frequency and flow to zero \cite {guinea3,herb} in the limit $\omega\rightarrow 0$, i.e., $g_r=8\pi/\ln(\Lambda/\omega)$, where $\Lambda$ is an ultraviolet cutoff.  This $\omega$ dependence can be probed experimentally to identify the effect of Coulomb interactions.

  For finite frequency $\omega$, however, there are additional diagrams ( like one showed in Fig.2 ), which connect two electron loops by an interaction line, in analogy with the dirty wire case \cite{nagaev}. In Appendix B we show that
  the shot noise contribution of this diagram  for low frequency and $\omega> q$ is proportional the the small transmitted momentum $q$, i.e. vanishes as $q\rightarrow 0$. Therefore,  we can neglect the diagrams with two electron loops.
 \begin{figure}
\begin{center}
\includegraphics [width=0.4 \textwidth ]{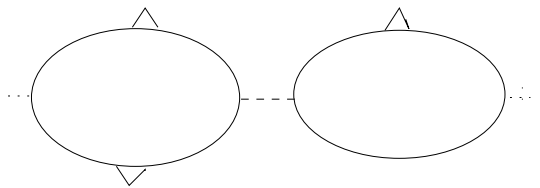}
\caption {The two loops diagram which describes the first order corrections to the noise. The lines are the same as in Fig.1. The  additional  diagrams of this kind  are those which replace the loop with two  Keldysh Green's function (like left one in Fig.2) by  retarded and advanced Green's functions.}
\end{center}
\end{figure}

 The renormalized $g_r$ is finite at $\omega=0$ if weak disorder generated by ripples is present. An attractive line of fix points appears then in the flow diagram \cite {guinea3,herb}
leading to a finite term in the interacting part of the noise even at $\omega=0$; however, other disorder corrections to the noise may arise, which are not addressed here.

Next we consider the implication of Coulomb interaction in graphene on the zero frequency shot noise, i.e.,  $V\gg T$. We take the limit $T\rightarrow 0$ and study the case of small voltages (we keep only terms linear in $V$).
 We substitute  the representation (\ref{GK}) for Keldysh GF and the solution (\ref{so2})  into  Eqs. (\ref{sv1}) and (\ref{se1}). We find a common factor for both the self energy and for the vertex part of the noise, which at $T\rightarrow0$   coincides with the term in the square brackets of Eq.(\ref{sn0r}) (we note that an integration which involves only one Keldysh Green's function results in $F_d\rightarrow1$ at $T\rightarrow0$). This factor replaces the value $1-f_0^2$ in the expressions for $I_2(p)$  (\ref{I2}) and in $J_1(q)$ (Eq. \ref{J1}). Thus in standard units we have
\begin{eqnarray}
S_g&=&\alpha\frac{2e|V|}{3}0.19\sigma_{min}g_r
\label{SN}
\end{eqnarray}

To find the Fano factor we extract from the equilibrium noise the conductivity and, by this calculate the Poissonian noise:
$S_P=2eI=2e|V|\alpha\sigma_{min}(1+0.19g_r)$.
Comparing with the expression for the shot noise
\begin{eqnarray}
S&=&S_{n0}+S_g=\alpha\frac{2e|V|}{3}\sigma_{min}(1+0.19g_r)
\label{ST}
\end{eqnarray}
we find $F=S/S_P=1/3$. Thus the unscreened Coulomb interaction precisely cancels and the Fano factor is independent of $g$.

 The value of Fano factor and its independence on interaction (at the first order) is in striking resemblance with analogous behavior in dirty wires \cite{gefen} when the limit of large inelastic length is considered. This  similarity between clean and dirty systems is due to evanescent modes which define the current-current correlations in pure graphene at the Dirac point. The diffusive type equation for distribution function is common to graphene and to dirty wires.  This distribution function preserves the Fano factor to the first order in $g_r$. We remark that in other cases $F$ does depend on the interaction to 1st order, e.g. in the interacting spinless resonance model and in the Kondo problem \cite{golub}. The  Fano factor is also changed by strong interaction (in the so called hot electron limit) in dirty wires with a short relaxation length. However, this case is described by a quasiequilibrium distribution function and is beyond the applicability of perturbation theory \cite{nagaev}.
 In the case of  ballistic
graphene the hot electron limit is unlikely since the interaction flows to small value at small $\omega$.

A few technical remarks are in order, related to the numerical coefficient of the coupling $g$.
The exact value of this coefficient depends on the order in which $\omega$ and
the finite inverse lifetime $\delta$ are taken to zero. We take the dc limit $\omega\rightarrow0$ while
 keeping $\delta>0$  (see  discussion in Refs.\cite{mirlin,ludwig}).
 The factor $(1-f_0^2(\epsilon))$ in the relevant integrals at $T\rightarrow0 $ supports
  such a choice. For the self-energy diagram we  chose
  hard cutoff which restricts the fermion energy to the Dirac point \cite{herb}. In works that use the Kubo formula for the conductivity an additional regularization procedure for the self energy (Fig.1a) is necessary, a procedure that is a  the subject of debate  \cite{herb,mish,daniel,herb2}. These differences may cause our numerical coefficient $0.19$ to deviate from that of Ref. \onlinecite{daniel} while being closer, though bigger, than the result $0.081$ of Ref. \onlinecite{herb}.

 \section { Conclusion}
 We apply the standard Keldysh technique and kinetic equation approach to calculate the
 shot noise power of clean graphene at the Dirac point  for large aspect ratio $W>>L$. Considering e.g. a sample \cite{hakonen} with $W/L=24, L=200nm$, the limitation of being at the vicinity of the Dirac point implies that $V, \omega, T <v_F/L\approx 20meV$. We suggest that tunneling experiments, as done in dirty wires \cite{pothier}, can be done in this voltage range and can test the linear interpolation of Eq. (\ref{fermi}), i.e. a position dependence of a two-step energy distribution function.
We have also found an interaction correction for the shot noise, varying as $g\sim 1/\ln(\omega)$ (Eq. (\ref{ST})). This coulomb effect may be detected by the frequency dependence of either the conductivity or the shot noise. We note that $v_F/L$ is the ballistic flight time across the system, hence at $\omega\ll v_F/L$ the noninteracting term is expected to be $\omega$ independent \cite{mirlin}. Therefore an observed $\omega$ dependence in this range can identify the elusive Coulomb interaction. Since the same correction appears in the conductance and in the shot noise the Fano factor remains robust to interactions and persists being equal to 1/3.
\begin {acknowledgments}
We would like to thank I. F. Herbut, L. S. Levitov, A. D.
Mirlin and S. A. Gurvitz for stimulating discussions. This research was supported by THE ISRAEL SCIENCE FOUNDATION (Grant No. 1078/07).
\end{acknowledgments}

\appendix
\section {}
In this appendix we evaluate noise diagrams in Fig. 1.
The self-energy diagrams can be written explicitely as
\begin{eqnarray}
S_{\Sigma}(t,t')&=&\frac{e^2 i g\alpha}{2L}\sum_{\sigma }\int
Tr[G_{\sigma}({\bf z}t_1{\bf x}t)\gamma_x \tau_x G_{\sigma}({\bf x}t{\bf x}'t')\gamma_x\tau_x \nonumber\\
&&G_{\sigma}({\bf x}'t'{\bf y}t_1)\frac{1}{|{\bf z}-{\bf y}|}(\gamma_0 G_{\sigma}({\bf y}t_1{\bf z}t_1)\gamma_0\tau_x+\nonumber\\
&&\gamma_0 \tau_x G_{\sigma}({\bf y}t_1{\bf z}t_1)\gamma_0)]
 \label{se}
\end{eqnarray}
and the vertex part acquires a form
\begin{eqnarray}
S_{v}(t,t')&=&\frac{e^2 i g\alpha}{2L}\sum_{\sigma }\int
Tr[\gamma_0 G_{\sigma}({\bf y}t_1{\bf x}'t')\gamma_x \tau_x G_{\sigma}({\bf x}'t'{\bf z}t_1)\gamma_0\nonumber\\
&&\frac{1}{|{\bf z}-{\bf y}|}(\tau_x G_{\sigma}({\bf z}t_1{\bf x}t)\gamma_x \tau_x G_{\sigma}({\bf x}t{\bf y}t_1)+\nonumber\\
&&G_{\sigma}({\bf z}t_1{\bf x}t)\gamma_x\tau_xG_{\sigma}({\bf x}t{\bf y}t_1)\tau_x)]
 \label{sv}
\end{eqnarray}
where symbol $ \int$ denotes multiple integrations $\int=\int dR d^2r d^2 y d^2 z dt_1$.

The gauge transformation (\ref{U}) yields a $V(t-t')$ dependence for all Green's functions. We simplify the noise by adding or subtracting expressions such as the energy integrals of the products of GF:  $G^{A} G^{A}$ or  $G^{R} G^{R}$ (which is zero). Thus for the zero frequency contributions to the noise power (at $\omega\rightarrow0$ ) we get
\begin{eqnarray}
S_{v}&=&\frac{2e^2 i g\alpha}{L}
 tr [\gamma_0 (G^{K}\gamma_x  G^{R}+G^{A}\gamma_x  G^{K})\nonumber\\
&&\hat{D}\gamma_0( G^K\gamma_x  G^{K}- G_-\gamma_x  G_-)]
 \label{sv1}
\end{eqnarray}
\begin{eqnarray}
S_{\Sigma}&=&\frac{e^2 i g\alpha}{L}
tr\{[G^A\gamma_x  (G^{K}\gamma_x G^{K}-G_- \gamma_x G_-)+ \nonumber\\
&& (G^K\gamma_x  G^{K} -G_- \gamma_x G_-)\gamma_xG^R]\hat{D}G_s^K\}
 \label{se1}
 \end{eqnarray}
Here $tr$  includes all summation in Dirac space and integrations on space and time variables. We also denote  $G_-=G^R-G^A$ and index $s$ refers to the relation $G^K_s =\gamma_0G^K \gamma_0$. Also we use notation  $\hat{D}=1/|{\bf z}-{\bf y}|$
and have dropped a term
 proportional to $G_s^R+G_s^A$ in the self-energy contribution which is zero by energy integration.

 The interaction part to the equilibrium noise $S_g=S_{\Sigma}+S_v$ can be calculated by applying the fluctuation dissipation theorem.
 Thus we have
 \begin{eqnarray}
S_g&=& ie^2g\alpha \int\frac{d^2p}{2\pi}\frac{d^2q}{(2\pi)^2}\frac{tr[I_2(p)I_1(q)+2J_2(p)J_1(q)]}
{|{\bf p}-{\bf q}|}\nonumber
\end{eqnarray}
here the trace is taken over Dirac matrices. The functions $J_i(p)$ and $ I_i(p)$ include energy integration
 \begin{eqnarray}
I_1(q)&=&\int\frac{d q_0}{2\pi}f_0(q_0)G_{s-}({\bf q}q_0)\\
I_2(p)&=&-\int\frac{d p_0}{2\pi}(1-f_0^2(p_0))(G'\gamma_xG_{-}\gamma_xG_-+\nonumber\\
&&C_- \gamma_xG_{-}\gamma_x G')_{{\bf p} p_0}\label{I2} \\
J_1(q)&=&-\gamma_0\int\frac{d q_0}{2\pi}(1-f_0^2(q_0))(G_-\gamma_xG_{-})_{\vec{q}q_0}\label{J1}\\
J_2(p)&=&\gamma_0\int\frac{d p_0}{2\pi}f_0(p_0)(G_-\gamma_xG^R+
G^A \gamma_xG_{-})
\end{eqnarray}
here $G'=ReG^R$.

 A direct energy integration yields
\begin{eqnarray}
I_1(q)&=&-if_0(q){\bm \gamma}\cdot {\bf q}/q \label{i1} \\
I_2(p)&=& 16\pi T\Delta^2(p){\bm \gamma}\cdot {\bf p}/p^2 \label{i2}\\
 J_1(p)&=&-8\pi T\gamma_0 \gamma_x \Delta^2(p)\label{j1}\\
J_2(q)&=&\frac{if_0(q)}{2q^3}\gamma_0 ({\bm \gamma}\cdot {\bf q}\gamma_x{\bm \gamma}\cdot {\bf q}
+q^2\gamma_x)\label{j2}
\end{eqnarray}
 where $\Delta(p)=\delta/(\pi(\delta^2+p^2))$ and $q,p=|{\bf q}|,|{\bf p}|$.
Integrating over momenta in the formula for noise and collecting all contributions we note that logarithmic divergences are exactly compensated in the sum of two types of diagrams. Thus the Coulomb contribution to the equilibrium ($V=0$) noise becomes $S_g/4=0.19\alpha T \sigma_{min}g$. The  prefactor 0.19 cames from  numerical integration of the second term in $S_g$ and corresponds to the nondivergent part of vertex diagram (Fig.1b).

\section{}

In this section we estimate two loop  diagrams shown in Fig 2. The summations over the Keldysh indices leads actually, to the two sets of such diagrams: one is proportional to the sum of retarded and advanced Coulomb Green's functions, while the other is proportional to Keldysh Coulomb Green's function. As our principal approximation we consider unscreened coulomb interaction to first order in $g$, while $\omega$-dependent parts of Green's functions include random phase approximation  polarization and, therefore, are a higher order in $g$. Thus, the set of diagrams that involves the Keldysh Coulomb Green's function is the second order in coupling constant, i.e. beyond the 1st  order considered here. As to the former  diagrams,   the relevant contribution may be expected in the first order in $g$. However, it simple show, and we proved this by direct calculations,  that this set of diagram is proportional  the small transmitted momentum $q$ (in our case $q\rightarrow 0$) and does not contribute to the noise power. Indeed, in the limit of unscreened coulomb interaction $ (D^R+D^A)$$=1/|\vec{x}-\vec{y}|$ the contribution to the noise which originates from this set of diagrams  can be written as
 \begin{eqnarray}
S_{loop}(xt,x't')&=&\frac{e^2 i g\alpha}{2L}\int d\vec{y}d\vec{z} dt_1 P(xtyt_1)\frac{1}{|{\bf z}-{\bf y}|}Q(zt_1x't')\nonumber
\end{eqnarray}
where $ Q(zt_1x't')$ represents the current response to external field and $ P(xtyt_1)$ is the current-density correlator of noninteracting electrons,
\begin{eqnarray}
Q(zt_1x't')&=&\sum_{\sigma }Tr G_{\sigma}({\bf z}t_1{\bf x}'t')\gamma_x \tau_x G_{\sigma}({\bf x}'t'{\bf z}t_1)\gamma_0\nonumber\\
 P(xtyt_1)&=&\sum_{\sigma }Tr G_{\sigma}({\bf x}t{\bf y}t_1)\gamma_0 \tau_x G_{\sigma}({\bf y}t_1{\bf x}t)\gamma_x\tau_x\nonumber
\end{eqnarray}
In equilibrium for $S_{loop}$ we have
 \begin{eqnarray}
S_{loop}(\vec{q}\rightarrow0,\omega)&\sim&i g \sigma_{min}\alpha P(\vec{q}\omega)\frac{1}{|{\vec{q}}|}Q(\vec{q}\omega)\nonumber
\end{eqnarray}
where we also introduced the Fourier transform for all functions.
At zero temperature $T=0$ the current-density correlator $ P(\vec{q}\omega)$ and the function $ Q(\vec{q}\omega)$ can be easily calculated. In the limit of low frequency and $\omega> q$ we obtain
  \begin{eqnarray}
  P(\vec{q}\omega)&=&\frac{q_x}{\sqrt{1-q^2/\omega^2}}\simeq q_x\\
  Q(\vec{q}\omega)&\simeq&\frac{-4iq_x}{\pi\omega}P_{0}
\nonumber
\end{eqnarray}
here $q=|\vec{q}|$ and $P_{0}$ is the high momentum cutoff.
  Thus the shot noise contribution of this diagram  for low frequency and $\omega> q$ is proportional the the small transmitted momentum $q$, i.e. vanishes as $q\rightarrow 0$.

\end{document}